\documentclass[twocolumn,showpacs,preprintnumbers,aps,prl,lengthcheck]{revtex4-1}

\usepackage[english]{babel}
\usepackage{amssymb}
\usepackage{amsmath}
\usepackage{graphicx}
\usepackage{dcolumn}
\usepackage{bm}
\usepackage{hyperref}
\usepackage{makeidx}
\usepackage[usenames,dvipsnames]{color}

\makeindex

\begin{document}

\title{Operator product expansion coefficients of the 3D Ising model with a trapping potential}
\author{Gianluca Costagliola $^{1}$}
\affiliation{
$^1$ Dipartimento di Fisica, Universit\`a di Torino and INFN, Via P. Giuria 1,
10125, Torino,
Italy.\\
}
\date{\today}

\begin{abstract}

Recently the operator product expansion coefficients of the 3D Ising model universality class have been calculated by
studying via Monte Carlo simulation the two-point functions perturbed from the critical point with a relevant field. We show that this method can
be applied also when the perturbation is performed with a relevant field coupled to a non uniform potential acting as
a trap. This setting is described by the trap size scaling ansatz, that can be combined with the general framework of
the conformal perturbation in order to write down the correlators $<\sigma (\mathbf {r})\sigma(0)>$, $<\sigma (\mathbf
{r})\epsilon(0)>$ and  $<\epsilon (\mathbf {r})\epsilon(0)>$, from which the operator product expansion coefficients can be estimated. We
find $C^{\sigma}_{\sigma\epsilon}= 1.051(3)$ , in agreement with the results already known in the literature,
and $C^{\epsilon}_{\epsilon\epsilon}= 1.32 (15)$ , confirming and improving the previous estimate obtained in the
uniform perturbation case.

\end{abstract}

\maketitle

\section{Introduction}

In recent years approaches based on the bootstrap technique \cite{rat1}-\cite{elsh3}, within the Conformal Field Theory
(CFT) framework, allowed to improve the precision on universal numbers needed to characterize the n-point
correlation functions, such as scaling dimensions and Operator Product Algebra (OPE) coefficients. In particular new
results have been obtained for the 3D Ising model universality class \cite{rglioz}\cite{rrych}, whose scaling
dimensions were already known with very high precision from Monte Carlo (MC) simulations \cite{hasen}. However a
similar comparison between different methods is still difficult for the OPE coefficients due to the lack of MC results.

In the reference \cite{our} a method based on the conformal perturbation theory was implemented for calculating the OPE
coefficients from off-critical correlators. The method exploits the short distance expansion of the
perturbed two-point functions, written as a power series of the Wilson coefficients derivatives and the
one-point functions. This procedure was already known \cite{gm}-\cite{cas}, but it was never applied to
the three dimensional case. In \cite{our} the practical feasibility of the method was shown and in particular
it was used to extract the OPE coefficients of the 3D Ising model, setting the magnetic
field as perturbation, and the following results were found: $C^{\sigma}_{\sigma\epsilon}=1.07(3)$ and
$C^{\epsilon}_{\epsilon\epsilon}=1.45(30)$.

In order to improve the precision of these results, and also to extend this procedure to another kind of perturbation, 
we consider now the Ising model perturbed from the critical point with a non uniform confining potential coupled to the spin, 
obtaining for example this Hamiltonian:
\begin{align}
\mathcal{H} &= - J  \sum_{<i j>}  \sigma_{i} \sigma_{j} - H  \sum_{i} \sigma_{i} - \sum_{i} \sigma_{i} U (r_{i})
\label{traph} \\
&\mbox{with}\;\;\; U(r_{i}) = \left( \frac{ |\overrightarrow{r_{i}} | }{l} \right) ^{p} =  |\overrightarrow{r_{i}} |^{p}
\cdot v^{p} \label{trapot}   
\end{align}
Where $r_{i}$ is the position vector of the site $i$ with respect to the central site of the lattice, $p$ is the power
exponent of the trap and $l = v^{-1}$ is the typical length. Therefore we have a central symmetry potential growing
towards the edges. 

The aim of this paper is to show that the conformal perturbation theory can be applied also in presence of a confining
potential and that the OPE coefficients of a model, in our case the 3D Ising model, can be
obtained by studying the correlation functions modified by the trap.

There are many reasons to motivate this study: the behavior of the critical system in presence of a trap is well
described by the trap size scaling (TSS), a framework based on renormalization group arguments
that allows to obtain the scaling laws of the observables at the critical point as a function of the trap parameters. 
The TSS has been successfully tested in many works both for quantum and classical models \cite{trap1}-\cite{trapn}. 

Besides the presence of a trap is a common feature of an experimental setup in which the system has
to be confined into a limited region. There are experimental studies reproducing this situation for systems in
the U(1) universality class \cite{trapex1}-\cite{trapex3}. 

Finally from a purely theoretical point of view it is interesting to combine the trap size scaling ansatz with the
perturbations in CFT, aiming to find the corrections to the two-point functions, because it
shows the general applicability of the method, even in this case in which the translational symmetry is explicitly
broken.

Moreover there could be some advantages, as we will see the TSS ansatz introduces a further external
parameter, that in principle could be used to improve the signal of the perturbation terms that have to be
calculated in the MC simulations.

\section{3D trapped Ising model}

\subsection{Trap size scaling ansatz}

In order to write down the correlation functions of the 3D Ising model with a trapping potential, it is useful to summarize the
general framework of the TSS as established in the original reference \cite{trap1}.

From the renormalization group framework (RG) we can write the scaling form of the singular part of the free energy
density, near the critical point, with the usual notation but keeping into account a further scaling variable $u_{v}$
given by the trap:
\begin{equation}
F_{s}( u_{t}, u_{h}, u_{v}, x ) = b^{-d} F_{s}( \; u_{t}b^{y_{t}} \;, u_{h} b^{y_{h}}, \; u_{v} b^{y_{v}}, x b^{-1} )
\end{equation}
Where we inserted also the space position $x$ to stress that now the translational invariance is lost and the
observables are point dependent. Iterating the RG transformation as in the standard approach (near
the critical point we can assume $u_{v} \sim v = l^{-1}$), we obtain:
\begin{equation}\label{fren}
F_{s} = l^{- \theta d} \; f ( t l^{\theta y_{t}}, h l^{\theta y_{h}}, x l^{-\theta} )
\end{equation}
Where we defined $\theta = y_{v}^{-1}$, that is the characteristic trap exponent and can be deduced by means of scaling
arguments from the corresponding continuum theory. In fact if we consider the potential (\ref{trapot}) as a
perturbation coupled to the spin relevant field of a generic conformal action, we can write:
\begin{equation}\label{actrap}
S = S_{cft} + S_{trap} \;\;\; , \;\;\; S_{trap} = \int \sigma (\mathbf {r} ) U(\mathbf {r} ) d \mathbf{r}.
\end{equation}
If we perform a scale change with parameter $b$, we have $r \rightarrow b^{-1} r$, $v \rightarrow b^{y_{v}} v$, and for
the field $\sigma \rightarrow b^{\Delta_{\sigma}} \sigma $, where $\Delta_{\sigma}$ is the spin scaling dimension. Since
they must compensate within the (\ref{actrap}), we
deduce:
\begin{equation}
\frac{1}{y_{v}} \equiv \theta = \frac{p}{d-\Delta_{\sigma}+p } =   \frac{2p}{d+2-\eta + 2p}
\end{equation}
This expression depends only on the dimensionality of the system, the geometry of the
confining trap and the universality class of the model.

From the free energy density at the critical point ($t =0$, $h=0$), we can deduce the scaling behavior of the one-point
functions of the spin and the energy in the central point of the trap, respectively:
\begin{align}
& < \sigma_{0} >  \; = \; B_{\sigma} \; l^{-\theta \Delta_\sigma} \label{mtrap}\\
& < \epsilon_{0} > \;  =  \; B_{\epsilon} \; l^{-\theta \Delta_\epsilon } \label{etrap}
\end{align}
Where $B_{\sigma}$, $B_{\epsilon}$ are non universal constants and the label zero emphasizes that we are considering
only local observables, typically defined in the central point, since in general the mean values are point
dependent.

Moreover the correlation length $\xi$ exhibits a scaling law $\xi \sim l^{\theta}$ with the trap parameters.
Therefore the singularities of the observables at the critical point are smoothed, being the correlation length limited
by the trap. However the TSS has a precise and universal scaling behavior determined by the typical length and the
exponent $\theta$. For this reason we expect the TSS can be applied also in our case for the perturbed correlation
functions.

\subsection{Correlators in the 3D trapped Ising model}

We shortly summarize the method already applied in \cite{our}. The starting point is the OPE \cite{wils} written
for perturbed systems:
\begin{equation}\label{wilson}
< O_i ({\bf r} ) O_j ({\bf 0} ) >_h = \sum _{k} C_{i j}^k (r, h) < O_k ({\bf 0}) >_h
\end{equation}
where the $O_i$ are a complete set of operators in CFT, the $< \;..\;>_h$ is the
expectation value over the action perturbed  with a small relevant parameter $h$. The Wilson coefficients $C_{i j}^k
(r, h)$ can be expanded as a Taylor series with respect to the perturbation parameter in a regular way without infrared
divergences, so that the coefficients of the expansion are the derivatives calculated at $h=0$ \cite{gm}. In particular we note
that the OPE coefficients can be extracted from the first correction term of the critical correlator. Hence by
calculating via MC simulations the perturbed two-point functions of relevant operators and the one-point functions,
whose behavior is well known from the RG framework, the OPE coefficients can be deduced. For further details see the
references \cite{our}-\cite{cas}. 
For example for the spin-spin correlator with the magnetic field perturbation we have:
\begin{equation}\label{example}
<\sigma (r)\sigma(0)> = \frac{1}{r^{2\Delta_\sigma}} \left( \; C^1 _{\sigma \sigma} + 
C^{\epsilon}_{\sigma\sigma} r^{\Delta_\epsilon} < \epsilon (0) > + \; ... \right)
\end{equation}
In our case the trap perturbation breaks the translational symmetry, but it still leaves the rotational symmetry due
to the shape of the confining potential (\ref{trapot}). Hence if we fix one operator in the center of the trap and the
other one at radial distance $r$, the (\ref{example}) is still valid and the fusion rules between fields are the same of
the uniform case. Then substituting the one-point functions (\ref{mtrap}) and (\ref{etrap}), we see that the expansion
can be written as power series of the scaling variable $s = v r^{1/\theta} $, that can be deduced also from the
(\ref{fren}). So there is perfect analogy with the uniform magnetic field case, provided that the scaling exponent of the 
trap and the scaling variable of the TSS ansatz are substituted, and we obtain straightforwardly:
\begin{align}
&< \sigma_{r} \sigma_{0} > \; = \frac{1}{r^{2 \Delta \sigma }} \left( \; C_{\sigma \sigma}^{1} +
C_{\sigma \sigma}^{\epsilon} \; B_{\epsilon} \; s^{\theta \Delta_\epsilon} + O(s^{\theta\Delta_\sigma +1}) \;\; \right)
\label{tcor1} \\
&< \epsilon_{r} \epsilon_{0} > \; = \frac{1}{r^{2 \Delta \epsilon }} \left( \; C_{\epsilon \epsilon}^{1} +
C_{\epsilon \epsilon}^{\epsilon} \; B_{\epsilon} \; s^{\theta \Delta_\epsilon} + O(s^{\theta\Delta_\sigma +1})
\;\; \right) \label{tcor2} \\
&< \sigma_{r} \epsilon_{0} > \; = \frac{1}{r^{\Delta \sigma + \Delta \epsilon }} \left( \; C_{\sigma
\epsilon}^{\sigma} \; B_{\sigma} \; s^{\theta \Delta_\sigma} +O(s) \;\; \right) \label{tcor3}
\end{align}
As in the general case, the expansion series for the correlators, taken from the center up to the distance $r$, converges
for distances less than about one correlation length. The desired OPE coefficients are included in the leading perturbation term, 
so that we can extract them knowing the TSS behavior of the one-point functions. Moreover there are two free parameters, $v$ and
$\theta$, that can be adjusted to maximize the relevance of the first correction term of the expansion. Our
aim is to check the validity of this scenario via MC simulations and to extract the 3D Ising model OPE coefficients.

\section{Monte Carlo simulation}

\subsection{Simulation settings}

We perform the simulations on a cubic lattice of side $L$ with fixed boundary condition. The trap is centered in the
middle point of the cube. On the lattice we denote with $\sigma_{r_{i}}^{lat}$ the spin located at distance $r$ from the
center upon the axis $i$. Hence we define the following observables: the spin one-point function on the central site $<
\sigma_0^{lat} >$,  and the energy one-point function in the middle of the lattice, defined as $< \epsilon_{0}^{lat} >
\equiv < \sigma_{0}^{lat}\sigma_{1}^{lat} > - E_{cr}$, where $E_{cr}$ is the energy bulk contribution at the critical
point, that must be subtracted.

The correlation functions are calculated from the central site of the lattice up to the distance $r$ on the
central axis, averaging between the six possible directions. Also for correlators involving the energy the constant
bulk contribution has to be subtracted. Therefore we have these definitions:
\begin{align}
&G_{\sigma \sigma} (r) \equiv \frac{1}{6} < \sum_{i=1}^{3} \sigma_{0}^{lat} \; (\sigma_{r_i}^{lat} + \sigma_{-r_i}^{lat}
) > \nonumber \\
&G_{\epsilon \epsilon} (r) \equiv \frac{1}{6} < \sum_{i=1}^{3} \epsilon_{0}^{lat} \; (\epsilon_{r_i}^{lat} +
\epsilon_{-r_i}^{lat} ) > \nonumber \\
&G_{\sigma \epsilon } (r) \equiv \frac{1}{6} < \sum_{i=1}^{3} \epsilon_{0}^{lat} \; (\sigma_{r_i}^{lat} +
\sigma_{-r_i}^{lat} ) > \nonumber 
\end{align}

Usually the site energy is obtained by averaging the products with the neighbor spins in all the directions.
However now there is no more translational invariance, therefore, except for the central site, the links in
different directions have not the same energy. However in our case the trap acts as a perturbation, so that the energy
difference between two neighbor sites due to the broken translational symmetry is negligible. We numerically find that 
this assumption is correct within the precision of our simulations.

The same attention is needed for the spin-energy correlation function, since in principle with the exchanging of the
two operators actually we obtain different correlators. However we anticipate that, for the distances and the trap
lengths involved in our simulations, we do not observe differences. This confirms the validity of the selected
window of perturbation parameters, since in the theory the operators can be exchanged without differences. 

The constraints to select the most appropriate trap parameter $v$ and the exponent $p$ are the same as in the uniform
magnetic field case, i.e. having a sufficient large correlation length to sample the correlator and avoiding the finite
size effects. After some tests we find that a potential with power $p = 2$ is the best choice, in fact with a
smaller exponent the correlation length of the system is short while with larger $p$ we have to increase too much the
size of the lattice to avoid finite size effects. We have checked that a cubic lattice of side $L = 400$ is
large enough to prevent them within our current precision. 

This is a very large lattice in three dimensions, however since all the observables are closely sampled around the
central site of the lattice, we can increase the speed of the simulation by using the hierarchical upgrade. This method
can be used when the observable is local: instead of doing the Monte Carlo sweep on the whole lattice
before sampling the observable, the sweep is performed only on a box around the point of interest. Actually the
algorithm works by defining many boxes of increasing side around the central point, and then performing nested cycles of
sweeps on every box. The detailed description can be found for example in the reference \cite{hasenbox}. In this way we
can save time and increase the maximum lattice size. Moreover we sample correlators at different distances in
different simulations, so that our data are uncorrelated.

We fix the following constants to their known value: the energy bulk contribution $E_{cr}=0.3302022
(5)$ and the critical temperature $\beta_{c} = 0.22165462 (2) $ from reference \cite{hasen2}, the scaling dimension
$\Delta_\sigma = 0.51815 (2) $ and $\Delta_\epsilon = 1.41267 (13)$ from \cite{rrych}, with whom we have $\theta =
2/(5-\Delta_\sigma) \simeq 0.44624$. The uncertainty on these constants is negligible in our data set.

With these settings we find that the optimal range for the trap parameter is $ 0.85 \cdot 10^{-5} 
\lesssim v \lesssim 2 \cdot 10^{-5}$ for which the correlation length is around 18 lattice 
spacings. In these simulations we choose a
thermalization time of $10^{4}$ sweeps, with starting configuration of all spins up aligned, and we sample about
$10^{8}$ configurations for each correlation function.

The two-point functions at the critical point have to be normalized to one as in the continuum, since, for example, 
the spin-spin correlator on lattice is:
$$ < \sigma_{i}^{lat} \sigma_{j}^{lat} > = \frac{R_{\sigma}^{2}}{|r_{ij}|^{2\Delta_\sigma}} $$
Therefore the spin normalization is $\sigma^{lat} = R_{\sigma} \sigma$, and similarly for the energy $\epsilon^{lat}
= R_{\epsilon} \epsilon$. From these we can convert all the quantities from lattice to continuum units. 
For further details see \cite{our} and \cite{cas}.

The constants $R_{\epsilon}$, $R_{\sigma}$ are also important because they fix the leading order term of the
spin-spin and the energy-energy correlators; as we see in the reference \cite{our} they are the main source of 
systematic errors on the final estimate of the OPE coefficients. We fix $R_{\sigma} = 0.550 (4)$ as in \cite{our},
while for $R_{\epsilon}$ we have refined the result by finite size scaling study of the critical correlator, obtaining
$R_{\epsilon} = 0.2377 (9)$.

Finally we have to fix the non universal constants $B_{\sigma}$, $B_{\epsilon}$ of the formulas (\ref{tcor1})-(\ref{tcor3})
by studying the one-point functions. We sample these observables in the optimal $v$ range selected 
previously, so that we find the expected power law behavior of formulas (\ref{mtrap})-(\ref{etrap}) 
without scaling corrections. We obtain $B_{\epsilon}^{lat} = 1.66 (2) $ and $B_{\sigma}^{lat} = 
1.485 (2) $.

\subsection{Two-point functions}

\begin{figure}[h]
\centering
\includegraphics[scale=0.35]{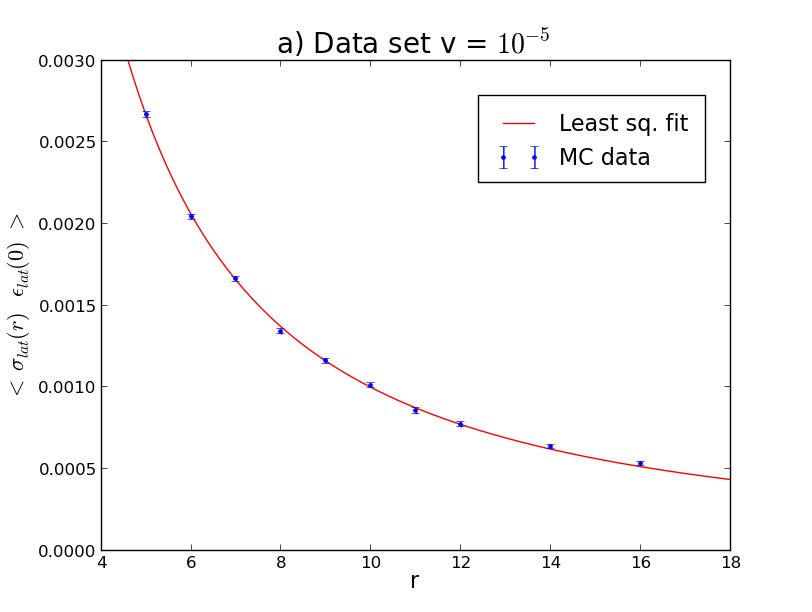}
\includegraphics[scale=0.35]{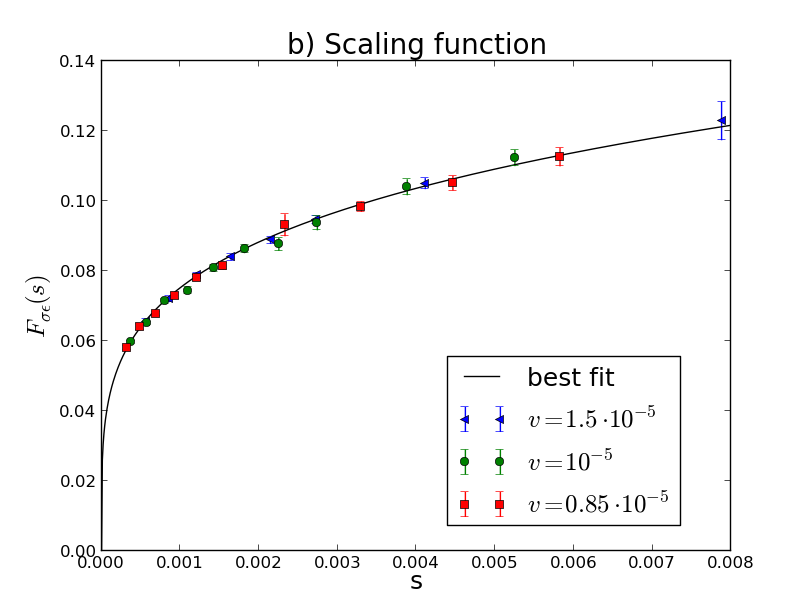}
\caption{\footnotesize a) Plot of the spin-energy correlation function data of the table \ref{tab_tem} for $v =
10^{-5}$. We can observe the good agreement between the MC data and the expected functional behavior of the perturbed
correlator. b) Plot of the scaling function $F_{\sigma \epsilon}$ for all the data series, showing they follow the
expected scaling behavior.}\label{fig_trapem}
\end{figure}

As we learn from the reference \cite{our}, the spin-energy correlator is the most suitable observable to determine the
coefficient $C_{\sigma\epsilon}^{\sigma}$ since it is zero at the critical point and there is not systematic error.
Hence we can fit only the leading order term of (\ref{tcor3}). The table \ref{tab_tem} shows the results of the fit
for the values of $v$ in the optimal range. The data sets are consistent with the functional form predicted by the TSS
ansatz. For comparison we remind the result of the bootstrap approach, for example from \cite{rrych},
$C_{\sigma\epsilon}^{\sigma} = 1.05184 (5) $. Our results are in good agreement with this estimate, and thanks to the
higher statistics we improve also the previous MC results. Combining all the data series we obtain
$C_{\sigma\epsilon}^{\sigma} = 1.051 (3)$. In the upper panel of figure \ref{fig_trapem} the data set for
$v=10^{-5}$ is shown, while in the lower panel all the rescaled data series are plotted in order to show that they are
distributed around the expected scaling function $F_{\sigma \epsilon}(s) \equiv <\sigma(r)
\epsilon(0)> r^{\Delta_\sigma+\Delta_\epsilon} = B_{\sigma}C_{\sigma\epsilon}^{\sigma} \; s^{\theta\Delta_\sigma}$.

\begin{figure}[h]
\centering
\includegraphics[scale=0.35]{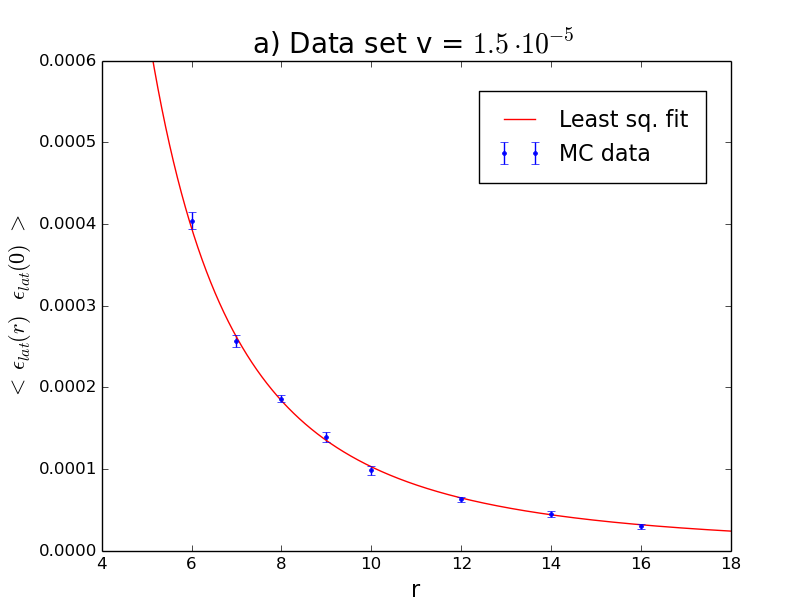}
\includegraphics[scale=0.35]{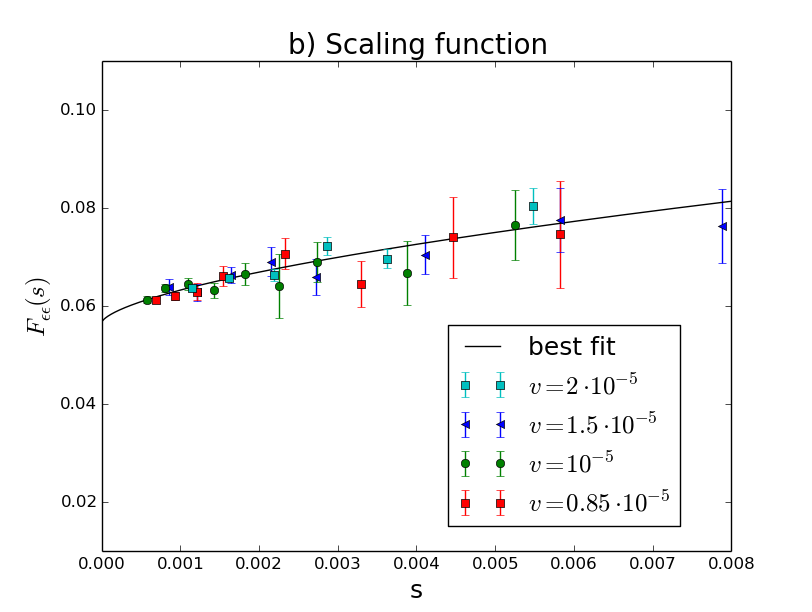}
\caption{\footnotesize a) Plot of the energy-energy correlation function data displayed in table \ref{tab_te} for
$v = 1.5\cdot10^{-5}$. b) Plot of the scaling function $F_{\epsilon \epsilon}$ for all the data series. Though this
observable has larger error bars than the spin-energy correlator, we can observe that the data are distributed
around the expected behavior.}\label{fig_trapee}
\end{figure}

The energy-energy correlator instead is the most interesting to determine the less known coefficient
$C_{\epsilon\epsilon}^{\epsilon}$. However this observable has a low signal due to the rapidly decreasing power law, and
it is affected by a systematic error due to the critical point term. We adopt the same convention of \cite{our} denoting the
fit statistical error in round brackets and the systematic error in square brackets. The table \ref{tab_te} shows our
results, that are fully consistent with the estimate of \cite{our}, and again we find a good agreement with the
expected power law behavior of (\ref{tcor2}). We quote the final estimate $C_{\epsilon\epsilon}^{\epsilon} =  1.32 (15)$. The
figure \ref{fig_trapee} shows the plot for the data set $v=1.5\cdot10^{-5}$ and the scaling function $F_{\epsilon
\epsilon}(s) \equiv <\epsilon(r)\epsilon(0)> r^{2\Delta_\epsilon} = R_{\epsilon}^{2} \;(
1+B_{\epsilon}C_{\epsilon\epsilon}^{\epsilon} \; s^{\theta\Delta_\epsilon} )$.

Finally we consider as consistency check the spin-spin correlator, since it is the observable with largest systematic
error. Moreover it is affected by short distance corrections, which instead were negligible for the other
correlators \cite{note}. The table \ref{tab_tm} shows the results, that are fully consistent with the previous estimates.

\begin{table}[h]
\centering
\begin{tabular}{|p{50pt}|c|c|c|c|}
\hline
 $v$     & $r_{min}$ & $r_{max}$ & \parbox[c]{50pt}{$ C_{\sigma \epsilon}^{\sigma} $} & \parbox[c]{40pt}{$
\chi^{2}/d.o.f.$}  \\
\hline
 $ 1.5 \cdot 10^{-5}$  & 5 & 16  & 1.054 (6)  & 0.7 \\
\hline
 $ 10^{-5}$            & 5 & 16  & 1.048 (7)  & 1.0 \\
\hline
 $ 0.85 \cdot 10^{-5}$ & 5 & 18  & 1.050 (7)  & 0.5 \\
\hline
\end{tabular}
\caption{\footnotesize Results of the one parameter least square fit for the spin-energy correlator for some values of
$v$. The columns $r_{min}$, $r_{max}$ report the range of sampled distances. The error quoted on $C_{\sigma
\epsilon}^{\sigma}$ takes into account also the standard error propagation from the normalization
constants.}\label{tab_tem}
\end{table}

\begin{table}[h]
\centering
\begin{tabular}{|p{50pt}|c|c|l|c|}
\hline
 $v$     & $r_{min}$ & $r_{max}$ & \parbox[c]{60pt}{$C_{\epsilon \epsilon}^{\epsilon}$} & \parbox[c]{40pt}{$
\chi^{2}/d.o.f.$}  \\
\hline
 $ 2 \cdot 10^{-5}$    & 6 & 12  & 1.34(5)[8]  & 1.1 \\
\hline
 $ 1.5 \cdot 10^{-5}$  & 6 & 16  & 1.32(11)[8]  & 0.5 \\
\hline
 $ 10^{-5}$            & 6 & 16  & 1.40(10)[10]  & 0.6 \\
\hline
 $ 0.85 \cdot 10^{-5}$ & 7 & 18  & 1.23(10)[10]   & 0.5 \\
\hline
\end{tabular}
\caption{\footnotesize Results of the one parameter least square fit for the energy-energy correlator for some values of
$v$. There are two error sources on $C_{\epsilon\epsilon}^{\epsilon}$: the statistical error (round brackets) takes into
account the fit uncertainty and the standard error propagation from the normalization constants, the systematic
error instead (square brackets) is due to the constant $R_{\epsilon}$ that fixes the non zero 
term of the correlator (\ref{tcor2}) at the critical point on the lattice.}\label{tab_te}
\end{table}

\begin{table}[h!]
\centering
\begin{tabular}{|p{50pt}|c|c|c|c|c|c|c|c|c|}
\hline
 $v$     & $r_{min}$ & $r_{max}$ & \parbox[c]{60pt}{$C_{\sigma \sigma}^{\epsilon}$} & \parbox[c]{40pt}{$
\chi^{2}/d.o.f.$}  \\
\hline
 $1.5 \cdot 10^{-5}$  & 5 & 16  &  1.057 (16) [50]  & 1.2 \\
\hline
 $10^{-5}$            & 5 & 16  &  1.048 (14) [60]  & 0.6 \\
\hline
$ 0.85 \cdot 10^{-5}$ & 5 & 18   & 1.061 (21) [70]  & 0.9 \\
\hline
\end{tabular}
\caption{\footnotesize Results of the two parameters fit for the spin-spin correlator with the same notations of the
table \ref{tab_te}. In this case the systematic error is large, actually more than the 
deviations of the observed values from the expected one, suggesting that the error bar of 
$R_{\sigma}$ is overestimated. However the results of the OPE coefficient are consistent with 
those of the spin-energy correlator.}
\label{tab_tm}
\end{table}

\section{Conclusions}

We have verified that the method of extracting the OPE coefficients from perturbed correlators works also in presence of
a trap: when the confining potential has a very large typical length compared to the distances involved for the
correlators, it can be considered as a perturbation and the OPE expansion for the perturbed two-point
functions can be wrote following the general prescription. 

The trap size scaling ansatz provides the tools to understand the behavior of the one-point functions and the exponents
of the power law terms, characterized by the trap exponent $\theta$. The possibility to combine these tools to
describe the perturbed correlators in principle was not ensured, due to the explicit translational symmetry breaking of
the potential coupled to the spin operator. Therefore finding the expected OPE coefficients of the 3D Ising
model universality class is itself a non trivial result. 

Moreover we can extract with a good precision the coefficients $C_{\sigma \epsilon}^{\sigma} = 1.051 (3) $ and
$C_{\epsilon \epsilon}^{\epsilon} = 1.32 (15)$; unfortunately the latter one, which is very interesting being the
most difficult to obtain also with the bootstrap approaches, has higher uncertainty than the former one.
However we are confident that this result could be improved with further MC simulations.

Another possibility could be coupling the confining potential to the energy operator, whose trap size scaling has
been equally investigated in \cite{trap1}. Given our results, there is no reason for which the same kind of
study could not be repeated in this other case. 

The main limitations to the trap approach are the large size of the lattice generally required (a problem partially
recovered by the hierarchical upgrades algorithm), and the fine tuning of the free parameters, the power exponent $p$
and the trap length $l$, which requires some preliminary studies to fix the non universal constants and to find the
most suitable sampling window. However this drawback is acceptable if it leads to isolate the terms
containing the OPE coefficients in the perturbed correlators expansion.

\vspace{20pt}
\textbf{Acknowledgments}
I would like to thank M. Caselle and N. Magnoli for useful suggestions and the INFN Pisa GRID data center for 
supporting the numerical simulations.

\begin{thebibliography}{}

\bibitem{rat1} R.~Rattazzi, V.~S.~Rychkov, E.~Tonni and A.~Vichi,
 JHEP {\bf 0812} (2008) 031 [arXiv:0807.0004 [hep-th]].

\bibitem{rych1} V.~S.~Rychkov and A.~Vichi,
Phys.\ Rev.\ D {\bf 80} (2009) 045006 [arXiv:0905.2211 [hep-th]].

\bibitem{rat2} R.~Rattazzi, S.~Rychkov and A.~Vichi,
Phys.\ Rev.\ D {\bf 83} (2011) 046011 [arXiv:1009.2725 [hep-th]].

\bibitem{pol} D.~Poland and D.~Simmons-Duffin,
JHEP {\bf 1105} (2011) 017 [arXiv:1009.2087 [hep-th]].

\bibitem{elsh1} S.~El-Showk, M.~F.~Paulos, D.~Poland, S.~Rychkov, D.~Simmons-Duffin and A.~Vichi,
Phys.\ Rev.\ D {\bf 86} (2012) 025022 [arXiv:1203.6064 [hep-th]].

\bibitem{papp} D.~Pappadopulo, S.~Rychkov, J.~Espin and R.~Rattazzi,
Phys.\ Rev.\ D {\bf 86} (2012) 105043 [arXiv:1208.6449 [hep-th]].

\bibitem{pol2} D.~Poland, D.~Simmons-Duffin and A.~Vichi,
 JHEP {\bf 1205} (2012) 110 [arXiv:1109.5176 [hep-th]].

\bibitem{elsh2} S.~El-Showk and M.~F.~Paulos,
Phys.\ Rev.\ Lett.\  {\bf 111} (2013) 241601 [arXiv:1211.2810 [hep-th]].

    
\bibitem{komar} Z.~Komargodski and A.~Zhiboedov,
 JHEP {\bf 1311} (2013) 140 [arXiv:1212.4103 [hep-th]]    
    
\bibitem{liendo} P.~Liendo, L.~Rastelli and B.~C.~van Rees,
JHEP {\bf 1307} (2013) 113 [arXiv:1210.4258 [hep-th]]   
    
\bibitem{glioz} F.~Gliozzi,
Phys.\ Rev.\ Lett.\  {\bf 111} (2013) 161602 [arXiv:1307.3111 [hep-th]]


\bibitem{kos} F.~Kos, D.~Poland and D.~Simmons-Duffin,
JHEP {\bf 1406} (2014) 091 [arXiv:1307.6856 [hep-th]]

\bibitem{kos2} F.~Kos, D.~Poland and D.~Simmons-Duffin,
JHEP {\bf 1411} (2014) 109 [arXiv:1406.4858 [hep-th]]     
    
\bibitem{gaiot} D.~Gaiotto, D.~Mazac and M.~F.~Paulos,
JHEP {\bf 1403} (2014) 100 [arXiv:1310.5078 [hep-th]]

\bibitem{elsh3} S.~El-Showk, M.~Paulos, D.~Poland, S.~Rychkov, D.~Simmons-Duffin and A.~Vichi,
Phys.\ Rev.\ Lett. {\bf 112} (2014) 141601 [arXiv:1309.5089 [hep-th]]



\bibitem{rglioz} F.~Gliozzi and A.~Rago,
JHEP {\bf 1410} (2014) 42 [arXiv:1403.6003 [hep-th]]

\bibitem{rrych} S.~El-Showk, M.~F.~Paulos, D.~Poland, S.~Rychkov, D.~Simmons-Duffin, A.~Vichi,
J.\ Stat.\ Phys. {\bf 157} (2014) 869  [arXiv:1403.4545 [hep-th]]

\bibitem{hasen} M.~Hasenbusch,
Phys.\ Rev.\ B {\bf 82} (2010) 174433 [arXiv:1004.4486  [cond-mat.stat-mech]]


\bibitem{our} M.~Caselle, G.~Costagliola, N.~Magnoli, 
Phys. Rev. D {\bf 91} (2015) 061901  [arXiv:1501.04065 [hep-th]]

\bibitem{gm} R.~Guida and N.~Magnoli,
Nucl.\ Phys.\ B {\bf 471} (1996) 361  [arXiv:hep-th/9511209]

\bibitem{gm2} R.~Guida, N.~Magnoli, 
Nucl.Phys. B {\bf 483} (1997) 563  [arXiv:hep-th/9606072]

\bibitem{gm3} R.~Guida, N.~Magnoli, 
Int.J.Mod.Phys. A {\bf 13} (1998) 1145 [arXiv:hep-th/9612154]

\bibitem{cas} M.~Caselle, P.~Grinza and N.~Magnoli, 
J. Phys. A {\bf 34} (2001) 8733 [arXiv:1501.04065 [hep-th]]


 


\bibitem{trap1} M. Campostrini, E. Vicari, 
Phys.\ Rev.\ Lett. {\bf 102}, 240601 (2009) [arXiv:0903.5153 [cond-mat.stat-mech]]

\bibitem{trap2} M. Campostrini, E. Vicari, 
Phys.\ Rev.\ A {\bf 81} (2010) 023606  [arXiv:0906.2640 [cond-mat.stat-mech]]

\bibitem{trapfs} S.L.A. de Queiroz, R.R dos Santos, R.B. Stinchcombe, 
Phys.\ Rev.\ E {\bf 81} (2010) 051122 [arXiv:1003.1075 [cond-mat.stat-mech]]

\bibitem{trap3} G. Costagliola, E. Vicari, 
J. Stat. Mech. L08001 (2011) [arXiv:1107.0815 [cond-mat.stat-mech]]

\bibitem{trap4} G. Ceccarelli, C. Torrero, E. Vicari, 
Phys.\ Rev.\ B {\bf 87} (2013) 024513  [arXiv:1211.6224 [cond-mat.stat-mech]]

\bibitem{trapn} G. Ceccarelli, J. Nespolo, 
Phys.\ Rev.\ B {\bf 89} (2014)  054504 [arXiv:1312.1235 [cond-mat.quant-gas]]

\bibitem{trapex1} F.M. Gasparini, M.O. Kimball, K.P. Mooney, M. Diaz-Avilla, 
Rev. Mod. Phys. 80, 1009 (2008)

\bibitem{trapex2} I. Bloch, J. Dalibarad, W. Zwerger, 
Rev. Mod. Phys. 80, 885 (2008)

\bibitem{trapex3} T. Donner, S. Ritter, T. Bourdel, A. Ottl, M. Kohl, T. Esslinger, 
Science 315, 1556 (2007) [arXiv:0704.1439 [cond-mat.stat-mech]]


\bibitem{wils} K.~G.~Wilson, 
Phys. Rev. {\bf 179} (1969) 1499 

\bibitem{hasenbox} M.~Caselle, M.~Hasenbusch, M.~Panero, 
JHEP {\bf 01} (2003) 057 [arXiv:hep-lat/0211012]

\bibitem{hasen2} M.~Hasenbusch, 
Phys. Rev. B {\bf 85} (2012) 174421 [arXiv:1202.6206 [cond-mat.stat-mech]]


\bibitem{note} We fit the spin-spin correlator with $< \sigma_{r} \sigma_{0} >_{lat} \; = R_{\sigma}^{2} r^{-2 \Delta \sigma} (1 +
{C_{\sigma \sigma}^{\epsilon}}^{lat} \; B_{\epsilon}^{lat} \; s^{\theta \Delta_\epsilon} ) + D r^{-(2 \Delta \sigma+1)}$, with 
${C_{\sigma \sigma}^{\epsilon}}^{lat}$ and $D$ as free parameters. The last term takes into account the short distance corrections
induced by the next to the leading order term of the correlator at the critical point on the 
lattice. This anstaz is sufficient to fit the correlator as reported in table \ref{tab_tm}.

 
 
\end{thebibliography}
\end{document}